\documentclass[final,5p,times,twocolumn]{elsarticle}

\usepackage[noend]{algpseudocode}
\usepackage[english]{babel}
\usepackage{verbatim}
\usepackage[table,xcdraw]{xcolor}
\usepackage{graphicx}
\usepackage{subcaption} 
\usepackage{amssymb}
\usepackage{amsmath}
\usepackage{amsfonts}
\usepackage{mathrsfs}
\usepackage{mathtools}
\usepackage{float}
\usepackage{multirow}
\usepackage{tabularx}
\usepackage{array}
\usepackage{color}
\usepackage{leftidx}
\biboptions{numbers,sort&compress}

\usepackage{textgreek} \usepackage[colorlinks,citecolor=blue,urlcolor=blue,bookmarks=false,hypertexnames=true]{hyperref}
\usepackage{comment}
\usepackage{layouts}

\usepackage{booktabs}
\usepackage{caption}
\definecolor{MC}{rgb}{0.0, 0.5, 0.0}

\newcommand*{\colorboxed}{}
\def\colorboxed#1#{%
  \colorboxedAux{#1}%
}
\newcommand*{\colorboxedAux}[3]{%
  \begingroup
    \colorlet{cb@saved}{.}%
    \color#1{#2}%
    \boxed{%
      \color{cb@saved}%
      #3%
    }%
  \endgroup
}

\usepackage{dashbox}

\usepackage{tikz}
\usepackage{soul}

\usepackage{rotating}

\usepackage{lineno}

\newcommand{\RN}[1]{%
  \textup{\uppercase\expandafter{\romannumeral#1}}%
}

\journal{arXiv}
\definecolor{azure(colorwheel)}{rgb}{0.0, 0.5, 1.0}

\definecolor{mygreen}{rgb}{0.0, 0.5, 0.0}
\definecolor{darkraspberry}{rgb}{0.53, 0.15, 0.34}
\definecolor{bleudefrance}{rgb}{0.19, 0.55, 0.91}

\newcolumntype{C}[1]{>{\centering\arraybackslash}m{#1}}

\newcommand{\eg}{\textit{e}.\textit{g}. }
\newcommand{\ie}{\textit{i}.\textit{e}. }

\begin{document}

\begin{frontmatter}


\title{A comparative study of transformer models and recurrent neural networks for path-dependent composite materials}

\author[mymainaddress]{Petter Uvdal}
\author[mymainaddress]{Mohsen Mirkhalaf\corref{mycorrespondingauthor}}

\cortext[mycorrespondingauthor]{Email: mohsen.mirkhalaf@physics.gu.se}

\address[mymainaddress]{Department of Physics, University of Gothenburg, Origovägen 6B, 41296 Gothenburg, Sweden}

\begin{abstract}
Accurate modeling of Short Fiber Reinforced Composites (SFRCs) remains computationally expensive for full-field simulations. Data-driven surrogate models using Artificial Neural Networks (ANNs) have been proposed as an efficient alternative to numerical modeling, where Recurrent Neural Networks (RNNs) are increasingly being used for path-dependent multiscale modeling by predicting the homogenized response of a Representative Volume Element (RVE). However, recently, transformer models have been developed and they offer scalability and efficient parallelization, yet have not been systematically compared with RNNs in this field.
In this study, we perform a systematic comparison between RNNs and transformer models trained on sequences of homogenized response of SFRC RVEs. We study the effect on two types of hyperparameters, namely architectural hyperparameters (such as the number of GRU layers, hidden size, number of attention heads, and encoder blocks) and training hyperparameters (such as learning rate and batch size). Both sets of hyperparameters are tuned using Bayesian optimization. We then analyze scaling laws with respect to dataset size and inference accuracy in interpolation and extrapolation regimes.
\textcolor{black}{The results show that while transformer models remain competitive in terms of accuracy on large datasets, the RNNs demonstrate better accuracy on small datasets and show better extrapolation performance.} When trained on a scarce dataset, we report a Root Mean Square Error (RMSE) of 9.0 MPa for the RNN compared to 10.6 MPa for the transformer model. At a larger dataset size both ANN models reach test errors around 3.5 MPa (RMSE), however, the maximum error of the transformer model remains higher. Furthermore, under extrapolation, there is a clear difference, where the RNN remains accurate with an RMSE of 5.4 MPa, while the transformer model performs poorly having an RMSE of 23.6 MPa. \textcolor{black}{ On the other hand, the transformer model is 7 times faster at inference, requiring 0.5 ms per prediction compared to the 3.5 ms per prediction for the RNN model.}
\end{abstract}

\end{frontmatter}
\section{Introduction}
\label{Introduction}
%
Polymer-based Short Fiber Reinforced Composites (SFRCs) are widely used in automotive, aerospace, and consumer products \cite{Friedrich2013, Timmis2015, ARULPRASANNA2024}, where lightweight, complex, and sustainable structural materials are valued \cite{Timmis2015, ARULPRASANNA2024}. SFRCs can be manufactured at low cost, yet are suitable for components with complex 3D geometries through processes such as injection molding \cite{FU20001117, MORTAZAVIAN2015116, EFTEKHARI2016153} and additive manufacturing \cite{Zhang2024, NAWAFLEH2020101109, SHAFIGHFARD2021101728}. Furthermore, natural fibers are also emerging as a promising reinforcement option in SFRCs, since they have lower CO$_2$ emissions during manufacturing, are more compatible with a recyclable life cycle, and therefore serve as a sustainable alternative to synthetic fibers \cite{Hagnell2019, Maurya2023}. Despite these advantages, predictive modeling of their nonlinear, history-dependent elasto-plastic behavior remains challenging.

Multiscale predictive modeling of SFRCs remains limited by the high computational cost of full-field simulations \cite{Bargmann2018, Mirkhalaf2022}. Conventionally, Finite Element (FE) or Fast Fourier Transform (FFT) methods are used to capture the microscale material behavior of a Representative Volume Element (RVE) with high-fidelity \cite{Qi2015, Spahn2014, Schneider2016, Mirkhalaf2020}. The microstructural behavior depends on fibers’ orientation, fiber volume fraction, and material properties of the constituent \cite{Advani1987, Hoang2016, Harper2012}. \textcolor{black}{For multiscale modeling, each Gaussian integration point of the macroscale model requires homogenized response of a unique RVE. However, this FE$^{\mathrm{2}}$ workflow often becomes unfeasible for SFRCs due to the quadratic scaling of computational cost. Instead, mean-field approaches (\eg Mori–Tanaka \cite{Mori1973} and earlier approaches \cite{Eshelby1957, HASHIN1962335, HASHIN1963127, Hill1965, BUDIANSKY1965223}) are typically used as an alternative.}

To address the challenge of high computational cost, data-driven Artificial Neural Networks (ANNs) have been developed, providing near-instantaneous predictions of material behavior \cite{Wang2018, Mozaffar2019, Ghavamian2019, Wu2020, Bonatti2022, Friemann2023, Ghane2023, Liu2023, CHEN2024113093, MAHARANA2025111176}. For history-dependent behavior, the first Recurrent Neural Networks (RNNs) for modeling plasticity in composite materials were introduced in 2018-2019 by implementing a Long Short-Term Memory (LSTM) as the functional unit of the network \cite{Wang2018, Mozaffar2019, Ghavamian2019}. Subsequent work extended these approaches to include Gated Recurrent Unit (GRU) variants \cite{Wu2020, Friemann2023, Cheung2024_transfer}. In addition to plasticity, various other deep learning approaches have also been developed to predict material properties of 3D-RVEs \cite{Mentges2021, GAJEK2021113952, MAHARANA2025111176}. For modeling plasticity, RNNs have proven effective in capturing sequential dependencies but suffer from limitations in long-term memory (see self-consistent RNNs as a strategy to mitigate memory losses \cite{Bonatti2022}), are sensitive to dataset size \cite{Cheung2024}, and retain residual internal uncertainties after training \cite{UVDAL2025111983}.

\textcolor{black}{For related machine-learning applications, transformer models were developed to effectively handle large dataset sizes.} The model was originally introduced by Vaswani et al. \cite{Vaswani2017} for language processing and can capture long-range dependencies in sequences via self-attention, which is computationally effective due to parallel processing. Since then, transformer models have had a large impact, for example Generative Pre-trained Transformer (GPT) Large Language Models (LLMs) such as ChatGPT by OpenAI \cite{radford2018improving, radford2019language}, AlphaFold to accurately predict protein 3D structures \cite{ jumper2021}, and AlphaEvolve (previously AlphaTensor) which found a faster algorithm for matrix multiplications \cite{Fawzi2022, Romera-Paredes2024, novikov2025}. In computational mechanics, recent studies have adapted transformer models as an alternative to RNNs for history-dependent material behavior, \ie to model the homogenized response of 2D microstructures \cite{Pitz2024, Zhongbo2024}. Pitz and Pochiraju \cite{Pitz2024} developed a pre-trained transformer model for random and cyclic loadin. Zhongbo and Poh \cite{Zhongbo2024} developed a pre-trained transformer model within an FE² framework by chunking long sequences. Despite these advances, there remains a need for a direct, systematic comparison between RNNs and transformer models.

The aim of this study is therefore to perform a systematic comparison of RNN and transformer architectures for modeling the nonlinear elasto-plastic response of SFRCs. Previous studies have used the Taguchi method \cite{nikzad2024} and Bayesian Optimization (BO) \cite{white2020,Kandasamy2018} to tune neural network architectures. In this paper, we use BO to optimize both the ANN architecture hyperparameters and its training hyperparameters. \textcolor{black}{We then analyze scaling behavior with dataset size, inference performance, and extrapolation behavior.} The results provide insight into the relative strengths of RNNs and transformers, offering practical guidance for the development of data-driven surrogates for composites.

The remainder of the paper is structured as follows. Section \ref{Dataset} introduces the publicly available dataset \cite{Cheung2024_transfer}, outlines the  augmentation method (adopted from \cite{Cheung2024}), and the preparation of data for this study. Section \ref{architectures} illustrates the RNN and transformer neural network architectures. Section \ref{Training} describes the BO and training procedure. In section \ref{Results}, we report and discuss the results, \ie, the scaling behavior with dataset size, inference and extrapolation performance. Final concluding remarks are summarized in Section \ref{Conclusion}.

\section{Dataset and data augmentation} 
\label{Dataset}

\subsection{Full-field simulations}
\label{simulations}
High-fidelity full-field micromechanical simulations of SFRCs have been developed in previous work by Cheung and Mirkhalaf \cite{Cheung2024_transfer}. Each sample corresponds to an RVE subjected to a six-dimensional random strain path of small sequential strain increments, solved with either FE- or FFT-based solvers in DIGIMAT-FE to obtain homogenized stress response of the RVE. Orientation tensors are generated using Arvo's algorithm \cite{Arvo1992}. The strain steps are generated, combined and scaled following Friemann et al. \cite{Friemann2023}. \textcolor{black}{From this, we get strain and stress sequences that we use for training the ANNs.}

\begin{figure}[htbp]
    \centering
    \includegraphics[width=0.25\textwidth]{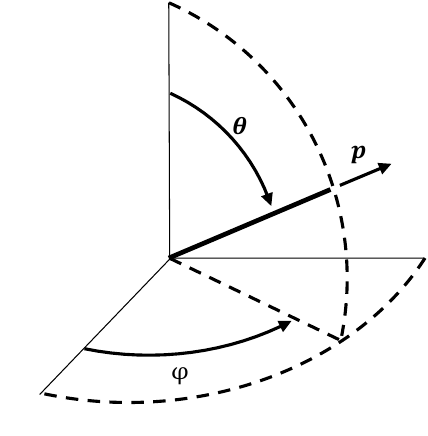}
    \caption{Fiber direction vector $\boldsymbol{p}$ expressed in spherical 
    coordinates $(\theta,\phi)$.}
    \label{fig:p_vector}
\end{figure}

Fiber orientation is compactly described by the second-order orientation tensor $\boldsymbol{a}$ \cite{Advani1987}, obtained from the fiber direction vector $\boldsymbol{p}$, as illustrated in Figure \ref{fig:p_vector}.
The fiber direction is parameterized in spherical coordinates $(\theta,\phi)$ as
\begin{equation}
\boldsymbol{p} = 
\begin{bmatrix}
\sin\theta\cos\phi \\
\sin\theta\sin\phi \\
\cos\theta
\end{bmatrix}, \qquad
a_{ij}=\int_{\Omega} p_i p_j \,\psi(\boldsymbol{p})\,\mathrm{d}\boldsymbol{p},
\end{equation}
where $\psi(\boldsymbol{p})$ is the orientation distribution function and 
$\Omega$ is the unit sphere.

The orientation tensor ($\boldsymbol{a}$) is symmetric with $\mathrm{tr}(\boldsymbol{a})=1$. Further details on the definition and properties of orientation tensors can be found in \cite{Advani1987, Cheung2024}.
Examples of RVEs and their corresponding orientation tensors are presented in Figure \ref{fig:rve}.
\begin{figure}[htbp]
    \centering
    \includegraphics[trim={0 0 0 0},clip]{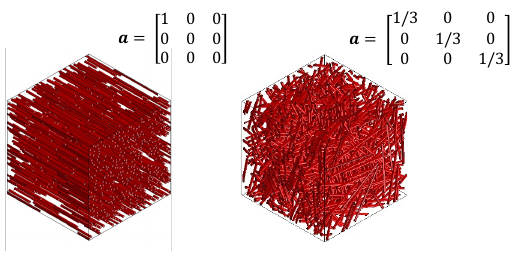}
    \caption{Example RVEs with their corresponding orientation tensor $\boldsymbol{a}$.}
    \label{fig:rve}
\end{figure}
The dataset contains 547 unique stress–strain sequences with varying fiber orientations and volume fractions, and has been made publicly available \cite{Cheung2024_transfer}.

\subsection{Data augmentation}
To address the inherent scarcity of high-fidelity simulations, Cheung et al. \cite{Cheung2024} proposed a rotation-based augmentation strategy. By applying random rotations $\boldsymbol{R}$ to all tensors, \ie the strain, stress and orientation tensors, additional training samples can be obtained by:
\begin{equation}
\boldsymbol{a}_{r} = \boldsymbol{R}\cdot\boldsymbol{a}\cdot\boldsymbol{R}^T, \quad
\boldsymbol{\epsilon}_{r}(t) = \boldsymbol{R}\cdot\boldsymbol{\epsilon}(t)\cdot\boldsymbol{R}^T, \quad
\boldsymbol{\sigma}_{r}(t) = \boldsymbol{R}\cdot\boldsymbol{\sigma}(t)\cdot\boldsymbol{R}^T,
\end{equation}
where $\boldsymbol{R}$ is a rotation tensor, obtained using Arvo's algorithm \cite{Arvo1992}. This preserves scalar invariants of the stress and strain tensors, while expressing the components in a rotated coordinate system. Rotation-based augmentation has shown to significantly improve performance of RNN models on scarce datasets \cite{Cheung2024}.

\subsection{Dataset preparation}
\label{Dataset prep}
For this study, we use the original dataset developed by Cheung and Mirkhalaf \cite{Cheung2024_transfer}, and divide the 547 samples into training (80\%), validation (15\%), and test (5\%) datasets. We and apply the rotation-based augmentation procedure of Cheung et al. \cite{Cheung2024} on the training and validation dataset, with augmentation index varying in the range of 1-20. The dataset contains 547 stress--strain sequences. We use 438 sequences for training, 83 for validation, and keep 26 sequences as a fixed test set. 
Rotation-based augmentation is applied only to the 438 training and 83 validation sequences, giving a total of 521 sequences available for augmentation. For an augmentation factor \(k\), the augmented dataset \(R_k\) contains \(N_k\) number of samples, where
\[
N_{k} = 521 \times k .
\]
For example, \(R1 = 521\) and \(R20 = 10,420\).

\section{Neural network architectures}
\label{architectures}

\subsection{Recurrent neural network model}
\label{RNN}
\begin{figure}[htbp]
    \centering
    \includegraphics[trim={0 6 0 6},clip]{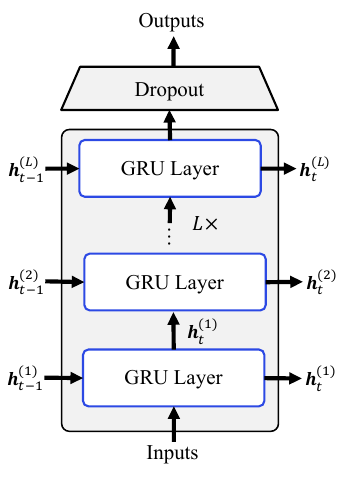}
    \caption{Schematic of a GRU-based recurrent neural network.}
    \label{fig:rnn}
\end{figure}

RNNs process sequential data by recurrently updating a hidden state vector $\boldsymbol{h}_t^{(l)}$ across timesteps in layer $l$, allowing temporal information to propagate through time. GRU-based architectures are particularly efficient for learning long sequences \cite{Cho2014}, and have been successfully applied to path-dependent plasticity in composites \cite{Friemann2023, Cheung2024, Ghane2023-RNN}. A schematic of the GRU-based RNN architecture used in this work is shown in Figure \ref{fig:rnn}.

\begin{figure}[htbp]
    \centering
    \includegraphics[width=0.45\textwidth]{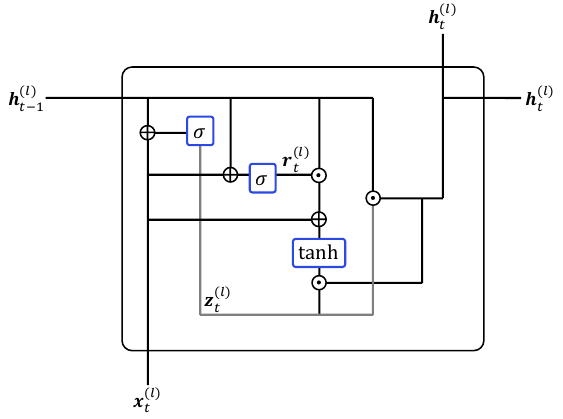}
    \caption{Schematic of the internal computation within a GRU cell, 
    showing update gate, reset gate, and candidate state.}
    \label{fig:gru_cell}
\end{figure}

Within a GRU cell in layer \(l\), the hidden state 
\(\boldsymbol{h}_t^{(l)}\) is updated using the update gate 
\(\boldsymbol{z}_t^{(l)}\) and reset gate \(\boldsymbol{r}_t^{(l)}\), 
as illustrated in Figure~\ref{fig:gru_cell}.

The update and reset gates of the GRU cell are computed as
\begin{align}
\boldsymbol{z}_t^{(l)} &= \sigma\!\left(W_z^{(l)} \boldsymbol{x}_t^{(l)} 
+ U_z^{(l)} \boldsymbol{h}_{t-1}^{(l)} + b_z^{(l)}\right), \\
\boldsymbol{r}_t^{(l)} &= \sigma\!\left(W_r^{(l)} \boldsymbol{x}_t^{(l)} 
+ U_r^{(l)} \boldsymbol{h}_{t-1}^{(l)} + b_r^{(l)}\right),
\end{align}
The candidate hidden state is then computed as
\begin{equation}
\tilde{\boldsymbol{h}}_t^{(l)} = 
\tanh\!\left(W_h^{(l)} \boldsymbol{x}_t^{(l)} 
+ U_h^{(l)} \big(\boldsymbol{r}_t^{(l)} \odot 
\boldsymbol{h}_{t-1}^{(l)}\big) + b_h^{(l)}\right),
\end{equation}
and the hidden state update is given by
\begin{equation}
\boldsymbol{h}_t^{(l)} =
(1 - \boldsymbol{z}_t^{(l)}) \odot \boldsymbol{h}_{t-1}^{(l)} 
+ \boldsymbol{z}_t^{(l)} \odot \tilde{\boldsymbol{h}}_t^{(l)}.
\end{equation}
The function \(\sigma\) is the sigmoid activation function,
\begin{equation}
\sigma(x) = \frac{1}{1 + e^{-x}}.
\end{equation}
and \(\odot\) denotes element-wise multiplication. Weight matrices 
\(W^{(l)}\) and \(U^{(l)}\) and biases \(b^{(l)}\) are trainable parameters.

In this work, after the GRU layers, a dropout layer with a probability of $0.5$ is applied to avoid overfitting, and the dropout probability is based on previous literature \cite{Cheung2024_transfer}. The stress prediction at time $t$ is then obtained from
\begin{equation}
\boldsymbol{\sigma}(t) = W_o \boldsymbol{h}^{(L)}_t + b_o,
\end{equation}
where $W_o$ and $b_o$ are the trainable output weights and biases, and $\boldsymbol{h}^{(L)}_t$ is the hidden state vector from the last GRU layer.

\subsection{Transformer model}
\label{Transformer}
\begin{figure}[htbp]
    \centering
    \includegraphics[trim={0 6 0 6},clip]{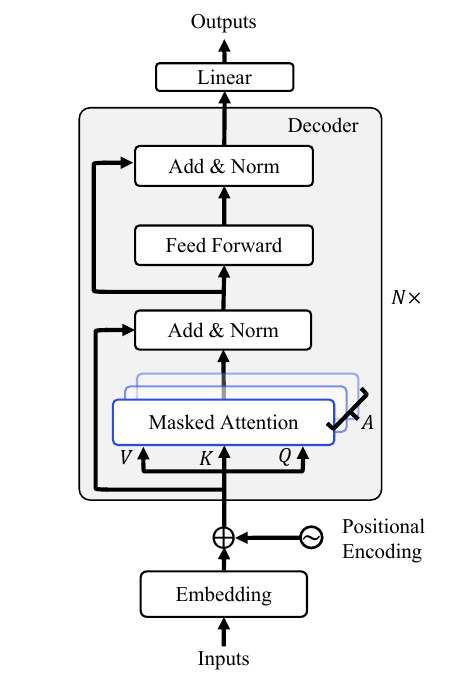}
    \caption{Schematic of the Transformer encoder architecture.}
    \label{fig:transformer}
\end{figure}

The transformer model replaces recurrence with multi-head self-attention, enabling the data from each timestep to attend to all others simultaneously \cite{Vaswani2017}. Input embeddings are combined with sinusoidal positional encodings to retain sequence order. Each encoder block consists of a multi-head attention layer followed by a feed-forward network, with residual connections and normalization \cite{Vaswani2017}. The overall architecture of a transformer encoder block is illustrated in Figure \ref{fig:transformer}.

To encode sequence order, sinusoidal positional encoding (PE) is added to the input embeddings:
\begin{equation}
\mathrm{PE}(t,2i)=\sin\!\Big(\tfrac{t}{10000^{2i/H}}\Big),\qquad
\mathrm{PE}(t,2i+1)=\cos\!\Big(\tfrac{t}{10000^{2i/H}}\Big),
\end{equation}
where $t$ is the timestep index, $i$ the feature dimension, and $H$ the model width. These terms are added element-wise to the input sequence $\mathbf{X}$ before attention.
Each input embedding is linearly projected to obtain the queries, keys, and values:
\begin{equation}
\mathbf{Q} = \mathbf{X}\mathbf{W}^Q,\qquad 
\mathbf{K} = \mathbf{X}\mathbf{W}^K,\qquad 
\mathbf{V} = \mathbf{X}\mathbf{W}^V.
\end{equation}
The scaled attention computes interactions between time steps:
\begin{equation}
\mathrm{Attention}(\mathbf{Q},\mathbf{K},\mathbf{V})
= \mathrm{Softmax}\left(\frac{\mathbf{Q}\mathbf{K}^{T}}{\sqrt{d_k}}+\mathbf{M}\right)\mathbf{V},
\end{equation}
where $d_k$ is the key dimension. The mask $\mathbf{M}$ has entries $0$ for allowed positions and $-\infty$ for masked positions (\ie future timesteps), ensuring that invalid connections receive zero probability after the row-wise softmax. Multi-head attention is obtained by repeating the attention operation with multiple learned projections $(\mathbf{W}^Q,\mathbf{W}^K,\mathbf{W}^V)$ and concatenating the results.

The output of the attention layer is combined with the original input through a residual connection and normalized. This is followed by a position-wise feed-forward network. Stacking several such encoder blocks enables the transformer to capture both short- and long-range dependencies in the sequence.

\section{Bayesian optimization and training}
\label{Training}
BO is used to automatically search the joint space of training hyperparameters (learning rate, batch size, number of epochs, etc.) and architectural hyperparameters (number of layers, hidden size, attention heads, feed-forward width, etc.). For each BO iteration, a set of hyperparameters is proposed by the acquisition function. The neural network is then trained from scratch using these hyperparameters on the original dataset, and its performance is evaluated on the validation set. BO uses this (validation-set) error to update its surrogate model of the hyperparameter landscape, and proposes a new set of hyperparameters for the next iteration. This sequential procedure is repeated and the hyperparameters that achieve the lowest validation RMSE are selected as the optimized architecture and training configuration. Once the optimized architectures and training hyperparameters are identified, we retrain the networks on datasets R1 through R20 (Section \ref{Dataset prep}) to evaluate scalability with increasing dataset size.

\subsection{Training and evaluation metrics}
The models are trained by minimizing the mean squared error (MSE) between the
predicted and reference stress components. The loss is defined as
\begin{equation}
\mathrm{MSE}
= \frac{1}{NTd} 
\sum_{n=1}^{N} \sum_{t=1}^{T} \sum_{i=1}^{d}
\left( \sigma_{i}^{(n)}(t) - \hat{\sigma}_{i}^{(n)}(t) \right)^{2}.
\end{equation}
where $\sigma_{i}^{(n)}(t)$ denotes the reference stress component for sequence
$n$ at timestep $t$, and $\hat{\sigma}_{i}^{(n)}(t)$ is the corresponding prediction produced by the neural network. The index $i$ runs over the $d$ independent stress components ($d=6$ for a symmetric second-order tensor in
Voigt notation), $N$ is the number of sequences, and $T$ is the sequence
length. During optimization and training, we also monitor the Root Mean Squared Error (RMSE), defined as the square root of the MSE.

To evaluate the model predictions, we use the von Mises equivalent stress, which provides a scalar measure that incorporates all six components of the stress tensor. The von Mises stress is physically meaningful for characterizing the material response, as it is directly related to yielding and effective stress in elasto-plastic materials. Using this metric therefore allows us to assess the accuracy of the network predictions in a way that is relevant to the underlying material response. For each timestep we calculate
\begin{equation}
\sigma_V = \sqrt{\tfrac{3}{2}\,\boldsymbol{\sigma}_{\mathrm{dev}} : \boldsymbol{\sigma}_{\mathrm{dev}} }, 
\qquad 
\boldsymbol{\sigma}_{\mathrm{dev}} = \boldsymbol{\sigma} - \tfrac{1}{3}\,\mathrm{tr}(\boldsymbol{\sigma})\,\mathbf{I},
\end{equation}
which is an invariant scalar under rotations and represents the effective stress. For a 
single sequence \(n\) of length \(T_n\), we define the Maximum absolute Error 
(MaE), the RMSE, the Mean Relative Error (MeRE), and the Maximum Relative Error 
(MaRE) as
\begin{align}
\mathrm{MaE}_{V}^{(n)} &= \max_{t} 
\left| \sigma_{V}^{(n)}(t) - \hat{\sigma}_{V}^{(n)}(t) \right|, \\[6pt]
\mathrm{RMSE}_{V}^{(n)} &= 
\sqrt{\frac{1}{T_n}\sum_{t=1}^{T_n} 
\left( \sigma_{V}^{(n)}(t) - \hat{\sigma}_{V}^{(n)}(t) \right)^2}, \\[6pt]
\mathrm{MeRE}_{V}^{(n)} &= 
\frac{\mathrm{RMSE}_{V}^{(n)}}{\max_{t} \sigma_{V}^{(n)}(t)}, \\[6pt]
\mathrm{MaRE}_{V}^{(n)} &= 
\frac{\mathrm{MaE}_{V}^{(n)}}{\max_{t} \sigma_{V}^{(n)}(t)} .
\end{align}

Here, the index \(n\) denotes the data sample (\ie one stress–strain 
sequence), and \(\max_{t}(\cdot)\) denotes the maximum over all timesteps of 
that sequence.

To report dataset-level performance, the sequence-level metrics are averaged 
over all \(N\) sequences in the dataset:
\begin{align}
\mathrm{RMSE}_{V} &= 
\frac{1}{N} \sum_{n=1}^{N} \mathrm{RMSE}_{V}^{(n)}, \\
\mathrm{MaE}_{V} &= 
\frac{1}{N} \sum_{n=1}^{N} \mathrm{MaE}_{V}^{(n)}, \\
\mathrm{MeRE}_{V} &= 
\frac{1}{N} \sum_{n=1}^{N} \mathrm{MeRE}_{V}^{(n)}, \\
\mathrm{MaRE}_{V} &= 
\frac{1}{N} \sum_{n=1}^{N} \mathrm{MaRE}_{V}^{(n)} .
\end{align}

\subsection{Bayesian optimization}
\label{bayesopt}
\begin{figure}[H]
    \includegraphics{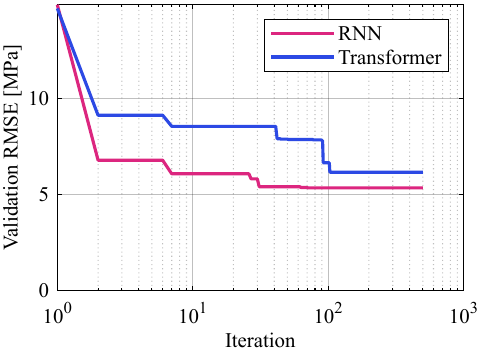}
    \caption{Bayesian optimization progress showing validation RMSE for RNN and Transformer models.}
    \label{bayes_compare}
\end{figure}

\begin{figure}[htbp]  \includegraphics{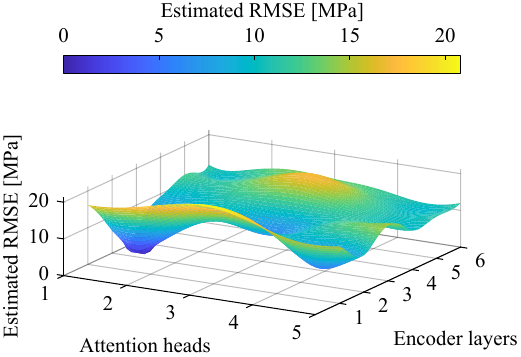}
    \caption{Validation RMSE as a function of attention heads and encoder layers for the transformer.}
    \label{bayes_surface_2}
\end{figure}

\begin{figure}[htbp]  \includegraphics{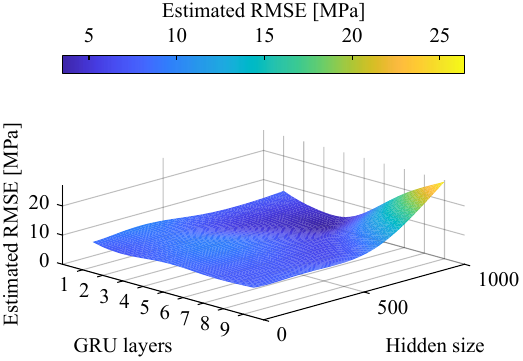}
    \caption{Validation RMSE as a function of GRU layers and the hidden size for the RNN.}
    \label{bayes_surface_3}
\end{figure}

\textcolor{black}{Bayesian optimization uses a Gaussian-mapping process}, balancing exploration and exploitation \cite{Snoek2012}. Each model is run with 200 trials using the BO, where the optimization progress is shown in Figure \ref{bayes_compare}.
While both models converged to low validation errors, the transformer achieved a minimum validation RMSE of 6.14 MPa, while the RNN reached a lower value of 5.33 MPa within fewer evaluations. \textcolor{black}{Furthermore, examples of optimized surface plots for neural network architectures hyperparameters are presented in Figures \ref{bayes_surface_2} and \ref{bayes_surface_3}. The figures show that transformer performance is sensitive to the number of attention heads and encoder layers, while the RNN becomes less accurate with a large hidden size and numerous hidden layers.}
For clarity, Table \ref{tab:hyperparams} summarizes all hyperparameters used to fully define the GRU-based RNN and transformer models, grouped into architectural and training hyperparameters. The listed ranges correspond to the search space explored by Bayesian optimization. For the RNN, the number of hidden layers ($L$) and the number of hidden states is also included. For, the transformer model, the number of masked attention heads ($A$) and decoder layers ($N$), and feed forward hidden size, are also optimized.
\begin{table}[h!]
\centering
\caption{Hyperparameters included in the Bayesian optimization search space for the RNN and Transformer models.}
\label{tab:hyperparams}
\small
\begin{tabular}{l c c}
\textbf{Hyperparameter} & \textbf{RNN (GRU)} & \textbf{Transformer} \\
\hline
\multicolumn{3}{l}{\textit{Architectural hyperparameters}} \\
\hline
Number of layers ($L$)                  
    & $[1,9]$                       
    & -- \\
Number of encoder blocks ($N$)           
    & --                                  
    & $[1,6]$ \\
Hidden size ($H$)                  
    & $[50,1000]$                  
    & $[50,500]$ \\
Feed-forward size          
    & --                                 
    & $[50,500]$ \\
Masked-Attention heads ($A$)                    
    & --                                 
    & $[1,5]$ \\
Dropout probability $p_{\mathrm{drop}}$
    & $0.5$
    & -- \\
\hline
\multicolumn{3}{l}{\textit{Training hyperparameters}} \\
\hline
Max epochs                                
    & $[100,800]$                         
    & $[100,800]$ \\
Mini-batch size                           
    & $[10,50]$                           
    & $[10,50]$ \\
Initial learning rate                    
    & $[10^{-4},10^{-3}]$                 
    & $[10^{-4},10^{-3}]$ \\
Learning-rate drop period ($\tau$)         
    & $[10,100]$                          
    & $[10,100]$ \\
Learning-rate drop factor ($\gamma$)        
    & $[0.9,0.99]$                       
    & $[0.9,0.99]$ \\
Gradient threshold                        
    & $[0.9,1.1]$                         
    & $[0.9,1.1]$ \\
\end{tabular}
\end{table}

The optimized architectures are summarized in Table \ref{tab:bo_results}. 
\begin{table}[htbp]
\centering
\caption{Optimized hyperparameters from Bayesian optimization for the RNN and Transformer models.}
\label{tab:bo_results}
\small
\begin{tabular}{l c c}
\textbf{Hyperparameter} & \textbf{RNN} & \textbf{Transformer} \\
\hline
Hidden size  & 817 & 56 \\
Feed-forward size & -- & 343 \\
Layers ($L$) / Encoder Blocks ($N$)     & 2   & 2 \\
Masked-Attention Heads ($A$)               & --  & 1 \\
\end{tabular}
\end{table}

\subsection{Training progress}
\label{bayesopt}
Figure \ref{fig:scaling} shows training and validation RMSE curves for the best RNN and Transformer models trained on the original dataset.
\begin{figure}[htbp]
    \includegraphics{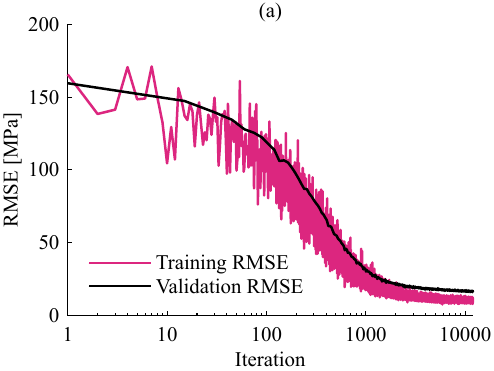}     \vspace{10pt} \\
    \includegraphics{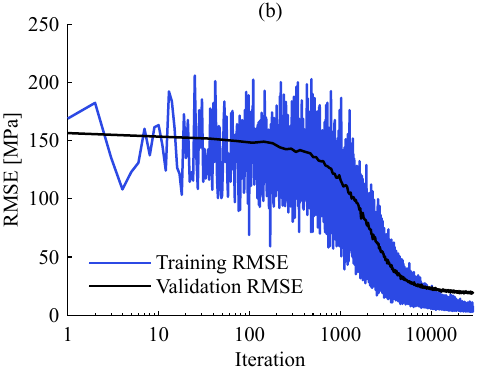}
    \caption{Training and validation RMSE for optimized (a) RNN and (b) Transformer.}
    \label{fig:scaling}
\end{figure}

After obtaining the optimized hyperparameters and network architectures, we retrain the models on the augmented datasets R1–R20 described in Section \ref{Dataset prep}. This enables us to systematically assess how model performance scales with dataset size. The results of this scalability study, together with inference and extrapolation tests, are presented and discussed in Section \ref{Results}.

\section{Results and discussion} 
\label{Results}
In this section, we present the main results of the study and evaluate the performance of the optimized RNN and transformer models. We first analyze how model accuracy scales with dataset size, followed by an assessment of inference behavior and extrapolation to cyclic loading paths.

\subsection{Scalability and dataset size}
\label{Inference}
We assessed different model performances under varying dataset sizes. For the test dataset (R0), the difference in performance is small, where the RNN had an RMSE of 8.96 MPa, while the corresponding value for the transformer model was 10.62 MPa. As the dataset size increased, the transformer models improved steadily, matching RNNs RMSE when trained on dataset R20. \textcolor{black}{The predicted error depending on augmentation index is presented in Figure \ref{fig:scaling}.}
\begin{figure}[htbp]
    \includegraphics{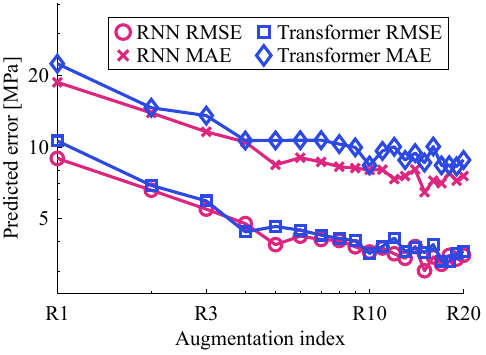}
    \caption{Scaling behavior with dataset size: test RMSE and MAE versus dataset size. RNN and transformer plotted with consistent color/marker coding across experiments.}
    \label{fig:scaling}
\end{figure}
The transformer model did, however, consistently have a higher MaE, which can be observed in Table \ref{tab:r20_rmse_mae_random}.

\begin{table}[h!]
\centering
\caption{MeRE and MaRE (relative errors, \%) and RMSE and MAE (MPa) for R20}
\label{tab:r20_rmse_mae_random}
\small
\begin{tabular}{l r r r r}
\textbf{Dataset (R20)} & \textbf{Test} & \textbf{Validation} & \textbf{Training} \\
\hline
RNN RMSE [MPa]         & 3.50 & 3.32 & 2.84 \\
RNN MaE [MPa]          & 7.56 & 7.04 & 5.99 \\
Transformer RMSE [MPa] & 3.64 & 3.45 & 1.98 \\
Transformer MaE [MPa]  & 8.83 & 7.65 & 4.38 \\
\hline
RNN MeRE               & 3.79\% & 3.53\% & 3.08\% \\
RNN MaRE               & 7.94\% & 7.46\% & 6.43\% \\
Transformer MeRE       & 4.01\% & 3.96\% & 2.26\% \\
Transformer MaRE       & 9.54\% & 8.94\% & 5.04\% \\
\end{tabular}
\end{table}

Moreover, the RNN model did show the highest accuracy on the test data with an RMSE of 3.01 MPa for R15, while the lowest RMSE of the transformer model was observed to be 3.31 MPa, when trained on R18.

These trends are consistent with prior observations that RNNs are effective in scarce-data regimes, while transformers benefit more from data scaling and parallel optimization \cite{Cheung2024,Zhongbo2024,Pitz2024}. Even though the RMSE for the transformer model is low on the test data (3.64 MPa), the higher MaE of the transformer model could suggest possible overfitting, or sensitivity due to the temporal encodings, especially since both the MaE and RMSE are significantly lower on the training data. In the next section, we explore the inference and extrapolation capabilities of the two networks.

\subsection{Inference and extrapolation}
\label{Inference}

\begin{figure}[htbp]
    \includegraphics{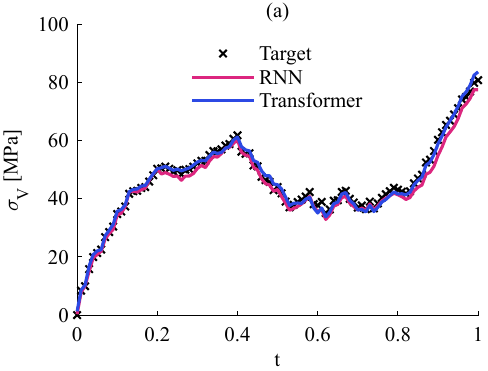}
    \vspace{2pt} %
    \includegraphics{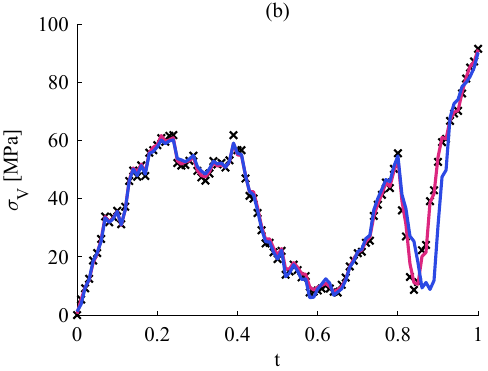}
    \caption{Network predictions against target values of von-Mises stress for two random multiaxial loading paths selected from the test dataset.}
    \label{fig:ab}
\end{figure}

For the validation dataset, both networks achieved their lowest RMSE and MaE when trained on the R20 dataset. We use the R20-trained models to evaluate accuracy on the independent test set. Figure \ref{fig:ab} shows two examples of von Mises stress predictions taken from two random strain paths in the test dataset. Although both models follow the reference response closely, the transformer performs slightly better for the first path (Figure \ref{fig:ab}a), whereas the RNN provides a more accurate for the second path (Figure \ref{fig:ab}b). These examples illustrate that, on random test sequences, neither architecture consistently dominates, and their relative performance depends on the specific loading path.

\begin{figure}[htbp]
    \includegraphics{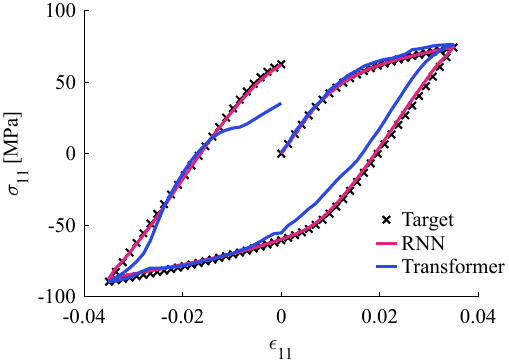}
    \caption{A uniaxial loading: micromechanical simulaiton against network predictions showing that RNN outperforms Transformer model in this case which is outside the training data.}
    \label{fig:c}
\end{figure}

\textcolor{black}{Extrapolation refers to evaluating the model on loading paths that differ from the training dataset, thereby testing its ability to generalize in other ranges or types of data. In the context of this paper, the RNN and Transformer models are trained exclusively on randomly generated multi-axial strain paths with maximum strain amplitudes in the range $0.01$--$0.05$. A mild extrapolation performance is then assessed using cyclic loading paths. We use standard cyclic tests, \eg uniaxial normal stress ($\sigma_{11}$), uniaxial shear stress ($\sigma_{12}$), biaxial stress in two normal directions ($\sigma_{11}{+}\sigma_{22}$), biaxial stress in normal and shear ($\sigma_{11}{+}\sigma_{23}$), and a plane strain ($\epsilon_{11}{+}\epsilon_{22}$). The cyclic loading starting at zero, goes up to a strain $0.035$, then to $-0.035$, and returns to zero strain. An example plot of a cyclic loading simulation, with the corresponding predictions of the ANNs, can be seen in Figure \ref{fig:c}.}
To further compare the models under the same five cyclic loading cases introduced above, we evaluate 11 RVEs under these loading paths. \textcolor{black}{Table \ref{tab:orientation_tensor_eigs} shows the fiber volume fraction and eigenvalues ($\lambda_{i}$) used for the RVEs.}
\begin{table}[ht]
\centering
\caption{Fiber volume fraction and eigenvalues ($\lambda_{i}$) of the orientation tensor $\mathbf{a}$ of RVEs simulated under the specific loading conditions.}
\label{tab:orientation_tensor_eigs}
\small
\begin{tabular}{c | c c c c}
\textbf{Sample} &$v_f$ & $\lambda_{1}$ & $\lambda_{2}$ & $\lambda_{3}$ \\
\hline
\textbf{1} & 0.115 & 0.79 & 0.20 & 0.02 \\
\textbf{2} & 0.132 & 0.60 & 0.35 & 0.05 \\
\textbf{3} & 0.141 & 0.89 & 0.08 & 0.03 \\
\textbf{4} & 0.114 & 0.79 & 0.17 & 0.04 \\
\textbf{5} & 0.131 & 0.65 & 0.28 & 0.07 \\
\textbf{6} & 0.125 & 0.54 & 0.27 & 0.19 \\
\textbf{7} & 0.111 & 0.46 & 0.42 & 0.12 \\
\textbf{8} & 0.129 & 0.86 & 0.12 & 0.02 \\
\textbf{9} & 0.119 & 0.55 & 0.42 & 0.03 \\
\textbf{10} & 0.148 & 0.96 & 0.04 & 0.00 \\
\textbf{11} & 0.125 & 0.35 & 0.33 & 0.32 \\
\end{tabular}
\end{table}
An example of the orientation tensor representation for a uniaxial test RVE is shown in Figure \ref{fig:rve}. The ellipsoid size correspond to the eigenvalues of the second-order orientation tensor $\mathbf{a}$, while $\mathbf{e}_1$, $\mathbf{e}_2$, and $\mathbf{e}_3$ denote the principal axes.
\begin{figure}[htbp]
    \centering
    \includegraphics{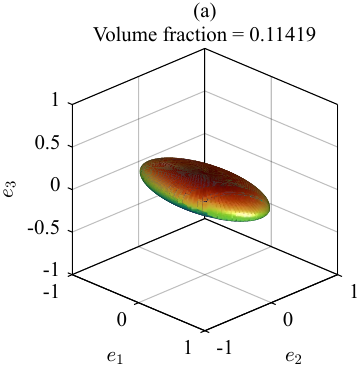}
    \includegraphics{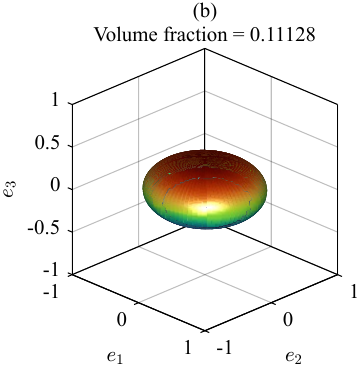}
    \caption{Orientation tensor ellipsoid for (a) sample 4 and (b) sample 7 used for the cyclic loading tests.}
    \label{fig:rve}
\end{figure}

\begin{figure*}[t]
    \centering    \includegraphics[width=0.8\textwidth]{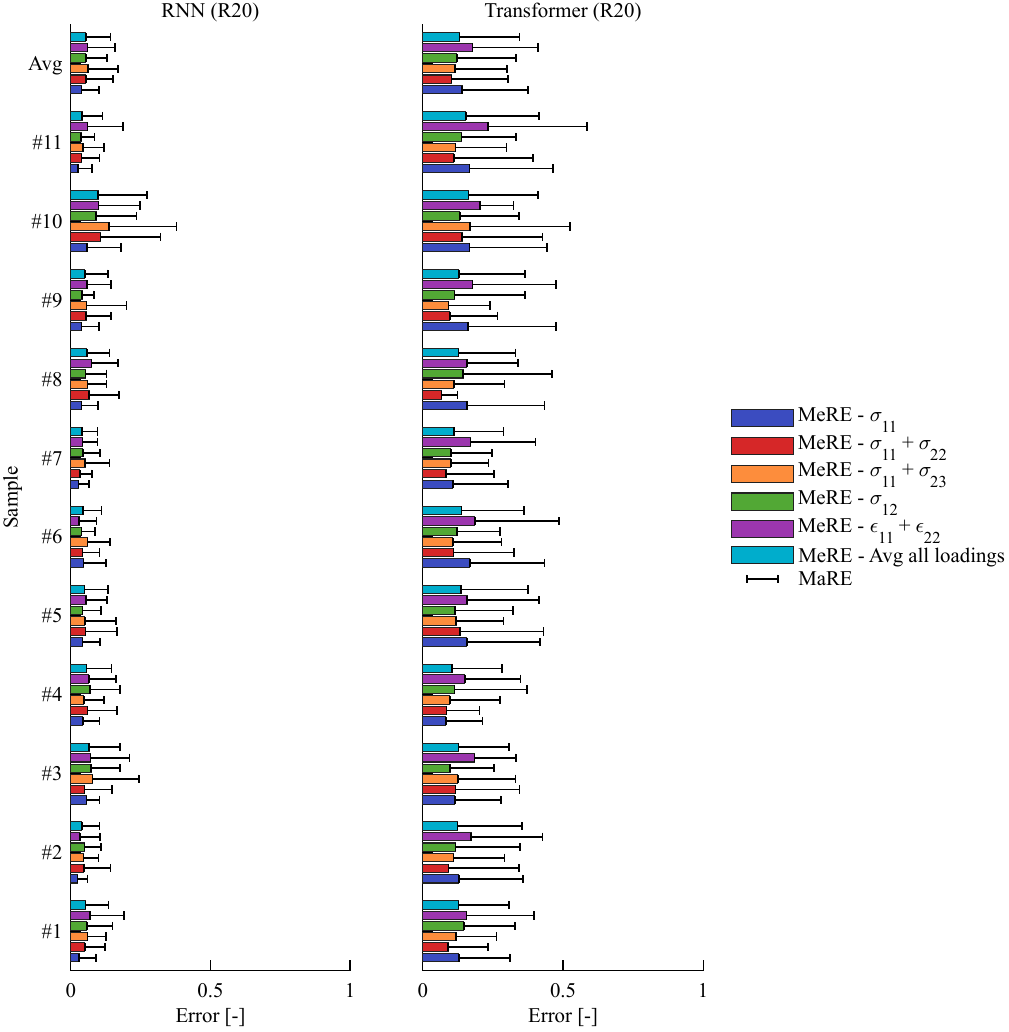}
    \caption{11 RVEs under the five cyclic loading paths: uniaxial normal stress ($\sigma_{11}$), uniaxial shear stress ($\sigma_{12}$), biaxial stress in two normal directions ($\sigma_{11}{+}\sigma_{22}$), biaxial stress in normal and shear ($\sigma_{11}{+}\sigma_{23}$), and a plane strain ($\epsilon_{11}{+}\epsilon_{22}$).
    Bars show the MeRE for each loading and error bars indicate the corresponding MaRE.}
    \label{fig:general_tests_barh}
\end{figure*}

\textcolor{black}{In extrapolation tests under cyclic loading, RNNs maintained stable predictions, while the transformer model did not manage to accurately capture cyclic loading behavior. The results are shown in Table \ref{tab:r20_rmse_mae_cyclic}.}
\begin{table}[ht]
\centering
\caption{The von Mises MeRE and MaRE (relative errors, \%) and RMSE and MAE (MPa), averaged for all cyclic loadings}
\label{tab:r20_rmse_mae_cyclic}
\small
\begin{tabular}{l c}
\textbf{Dataset} & \textbf{Cyclic loading} \\
\hline
RNN RMSE [MPa]          & 5.74  \\
RNN MaE [MPa]           & 15.09  \\
Transformer RMSE [MPa]  & 13.42  \\
Transformer MaE [MPa]   & 34.78  \\
\hline
RNN MeRE                & 5.37\%  \\
RNN MaRE                & 14.15\% \\
Transformer MeRE        & 13.07\%  \\
Transformer MaRE        & 34.46\% \\
\end{tabular}
\end{table}
For each sample, von Mises errors for these tests are summarized in Figure \ref{fig:general_tests_barh}.

Despite earlier reports of transformers’ superior ability to accurately capture complex temporal dependencies \cite{Vaswani2017}, we observed that they fail to extrapolate reliably under specific (cyclic) loading cases. This limitation may stem from sensitivity to temporal encodings or insufficient training data.

On the other hand, transformers evaluated sequences nearly seven times faster than RNNs (0.5 ms versus 3.5 ms per sequence on an NVIDIA RTX A4500) due to parallelization. Since surrogate ANNs are called at every Gauss integration point in coupled multiscale simulations, such a speedup could substantially reduce overall computational cost. In addition, their parallel structure makes transformers attractive for training on larger and more complex datasets with longer temporal dependencies.

\subsection{Further comparisons}

Beyond accuracy and inference speed, practical deployment in multiscale FE simulations introduces further considerations. For transformers, Zhongbo and Poh \cite{Zhongbo2024} proposed a chunking strategy to divide long load sequences, which alleviates memory demands of the CPU. \textcolor{black}{RNNs have difficulties resolving long temporal dependencies, since repeated updating of the hidden states eventually can overwrite information stored form the initial timesteps \cite{Liu2023}. Furthermore RNNs may be difficult to implement in multiscale FE-frameworks, and special care is required to access hidden state vector \cite{LOGARZO2021113482, Zhongbo2024}.} This shows that additional challenges still exist for RNNs, whereas transformers mainly face challenges related to data scarcity and robustness.

Overall, our results show that RNNs remain advantageous in scarce-data regimes and extrapolation tasks, here referring specifically to cyclic loading sequences. Transformers require more data and careful tuning, but provide clear benefits in inference speed and scalability. These findings align with recent trends in surrogate modeling for composites \cite{Zhongbo2024, Pitz2024}, suggesting that the most suitable architecture depends on the intended application, and data availability.


\section{Conclusions}
\label{Conclusion}

To the authors' knowledge, this study presented the first systematic comparison of RNN and transformer architectures for path-dependent materials. Bayesian optimization (BO) was used to ensure a fair and reproducible selection of both network architectures and training hyperparameters, avoiding manual tuning and suboptimal designs. The main findings are:

\begin{itemize}
\item RNNs achieved lower errors in scarce-data regimes and provided more reliable extrapolation to cyclic loading sequences.
\item Transformers reached comparable accuracy to RNNs when trained on sufficient dataset size, offered faster inference, and can be scaled more efficiently during training due to parallelized data processing.
\end{itemize}

These results provide practical guidance for choosing suitable architectures for surrogate modeling of path-dependent composite materials, depending on data availability and intended application. Although demonstrated here for short fiber reinforced composites, we expect that the comparative trends observed between RNNs and transformer models also apply to other classes of path-dependent materials. Future work could explore physics-informed or physics-encoded networks, hybrid RNN–transformer approaches, and uncertainty quantification techniques such as Monte Carlo dropout.

\section*{Acknowledgment}
The computations were enabled by resources provided by Chalmers e-Commons at Chalmers University of Technology.

\bibliographystyle{unsrt}
\bibliography{2.References}

\end{document}